\documentclass[slac_one]{revtex4}
\usepackage{graphicx}
\usepackage{fancyhdr}
\begin{document}

\title{Status of the T2K 280m Near Detector} 

%

\author{Thomas Lindner, on behalf of the T2K-ND280 Group}
\affiliation{University of British Columbia, Vancouver, Canada, V6T 1Z1}

\begin{abstract}
 
The Tokai-to-Kamioka (T2K) long-baseline neutrino-oscillation experiment
is being prepared for start of operations in Fall 2009. The 
purpose of T2K 
is to measure the oscillation parameters from an intense $\nu_{\mu}$ beam. In
order to make precision measurements of $\nu_{\mu}$ disappearence
a new detector complex (ND280) will be used to
identify the profile and composition of the neutrino beam near its
production site. ND280 will also be used to study important background processes
in the measurement of a $\nu_{e}$ appearance signal at Super-K. We will 
describe the physics goals and technology choices of ND280.
\end{abstract}

\maketitle

\thispagestyle{fancy}


\section{INTRODUCTION} 

T2K is a long-baseline neutrino oscillation experiment based in Japan.
The goal of T2K is to precisely measure the oscillation parameters associated 
with $\nu_{\mu}$ disappearance and search for evidence of 
$\nu_{e}$ appearance.  T2K will use a narrow-band off-axis beam of $\nu_{\mu}$
generated from the 50 GeV proton synchrotron of the new JPARC facility.
The beam will be pointed 2.5$^{o}$ off-axis from the upgraded
Super-Kamiokande detector. Using an off-axis beam results in a neutrino energy
spectrum that is narrowly peaked at 600-700 MeV, which is 
near the oscillation maximum for the 295 km baseline from JPARC to Super-Kamiokande.

With an integrated beam intensity of $5 \times 10^{21}$ protons on target, 
the T2K experiment will be capable of determining the oscillation parameters relevant
to the muon neutrino disappearance with a high precision of  $\approx 10^{-4}$ eV$^{2}$/c$^{4}$ 
and "¡­ $10^{-2}$ for $m^{2}_{23}$
and $\sin^{2} 2\theta_{23}$, respectively. 
The measurement of $\nu_{e}$
appearance in the same data sample is expected to 
improve the sensitivity to $\sin^{2} 2\theta_{13}$
by an order of magnitude compared to the current upper limit, 
reaching $\approx$ 0.008 for CP
violating phase $\delta_{CP}$ = 0.

The oscillation parameter measurements will be derived from measurements of 
the rates and spectrum of $\nu_{\mu}$ and $\nu_{e}$ seen in Super-Kamiokande.
However, in order to reach the design sensitivity, we must make 
precise measurements of the neutrino flux and spectrum before any oscillations have
occurred, ie measurements near the production site.  To this end we are 
building a new detector facility, called the Near Detector at 280m (ND280 
for short). This facility 
is located at JPARC, at 280 m from the beam target.
In what follows we will introduce the ND280 facility, as well as 
describing the physics goals and detector technologies choices of ND280.  
We will conclude with the current status and near-term 
future of ND280.

\section{ND280}



\begin{figure*}[t]
\centering
\includegraphics[width=100mm]{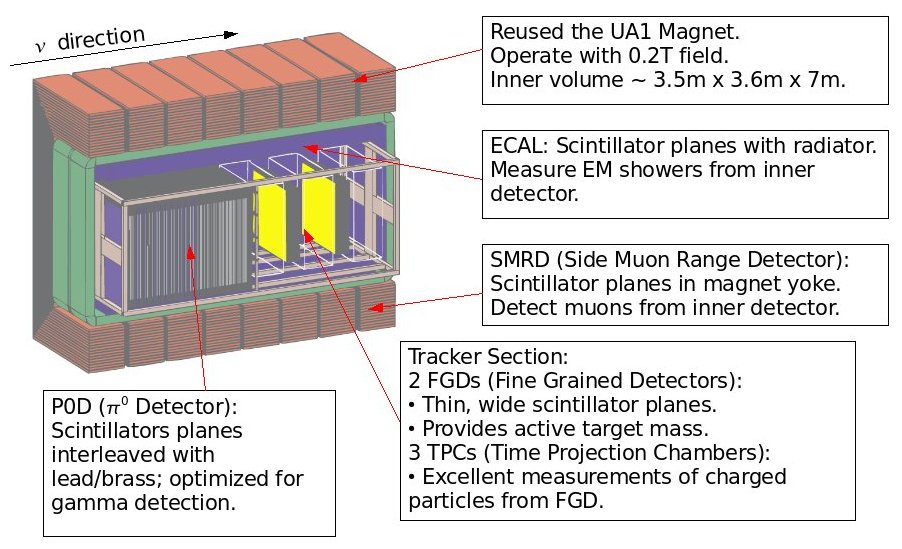}
\caption{Diagram of the ND280 Off-axis Detector.} \label{nd280_diagram}
\end{figure*}

There are  two different neutrino detectors in the ND280 pit.  
The first of these is INGRID, which is a scintillator-based detector that 
is centred on the axis of the neutrino beam.  This detector is designed to provide rapid feed-back
to the beamline operations group concerning the flux and position of the neutrino beam.

The second detector is the ND280 off-axis detector, which sits in the 
direction of Super-Kamiokande, as seen from the average neutrino production point.
Figure \ref{nd280_diagram} shows the off-axis detector in more detail.
We have re-used the UA1 magnet, which has been refurbished and shipped to Japan.
As seen from the diagram, there are five different detection sub-systems in
this detector; a set of Time Projection Chambers and four scintillator based 
sub-detectors.

\subsection{Physics Goals of ND280}

In order for T2K to succeed, ND280 must provide the following physics input:
\begin{itemize}
\item Measurement of unoscillated $\nu_{\mu}$ flux and spectrum, both on-axis 
and off-axis (towards Super-Kamiokande).
\item Measurement of small  $\nu_{e}$ flux that is intrinsic 
to beam (0.5\% $\nu_{\mu}$ flux).
\item Measurement of differential rates and characteristics of various 
neutrino interaction modes.
\end{itemize}

As noted above, the ND280 facility consists of a number of different sub-systems.  Different
parts of the detector have been optimized in order to make different measurements.  
Many of the measurements can be made in a complementary fashion with different parts of
ND280.
In what follows we shall explain two of the different measurements that will be 
made with ND280.  
These two examples do not cover the totality of what ND280 can achieve; rather they are
merely to give a sense of the measurements that will be performed.

\subsubsection{CCQE $\nu_{\mu}$ Interactions}

At the peak energy of the T2K off-axis beam the most probable interaction mode
for $\nu_{\mu}$ is the Charge Current Quasi-Elastic (CCQE) interaction:
\begin{equation}\label{eq_ccqe}
\nu_{\mu} + n \rightarrow  \mu^{-} + p.
\end{equation}
This is a clean interaction mode to reconstruct, since the initial 
neutrino energy can be approximately 
reconstructed from the muon alone.  This is the 
interaction mode that Super-Kamiokande uses for measuring the $\nu_{\mu}$
flux and spectrum.  

The Tracker section of the ND280 detector consists of the three TPCs 
interleaved with the two FGDs.  The Tracker has been optimized for making 
measurements of CCQE interactions.  The FGDs, which provide the target
mass for the interactions, are narrow in the beam direction, which means 
that a large fraction of the charged products from CCQE interactions 
enter the TPCs.  The TPCs will provide excellent tracking 
and particle identification for charged tracks that enter it and the FGD
will provide decent particle identification for shorter tracks that
don't enter the TPC (such as lower energy protons). The Tracker
is therefore well-adapted to measuring all the products of CCQE 
interactions.
In addition, the surrounding ECAL will provide information that will allow 
the rejection of events that might mimic CCQE interactions.

Note that one FGD is all plastic scintillator, but the other FGD interleaves 
plastic scintillator planes with water planes.  The partial-water FGD 
will allow us to extract the CCQE reaction rates on water, which is important
since Super-Kamiokande is a water-based detector.

\subsubsection{Neutral Current $\pi^{0}$ $\nu_{\mu}$ Interactions}

Another important interaction mode is the Neutral Current $\pi^{0}$ interaction:

\begin{equation}\label{eq_ncpi0}
\nu_{\mu} + N \rightarrow   \nu_{\mu} +\pi^{0} + N.
\end{equation}

Measuremement of the rate of this interaction is crucial, 
because this mode is an important
background for $\nu_{e}$ measurements at Super-Kamiokande 
(if one gamma from $\pi^{0}$ is missed by Super-Kamiokande,
then the event looks like a $\nu_{e}$ interaction).
We must therefore carefully measure the rate and characteristics
of this interaction.
The P0D detector has large target mass and lead radiators,
which allows it to efficiently reconstruct EM showers from 
$\pi^{0}$.  The P0D in combination with the surrounding 
ECAL will therefore be capable of making clean 
measurements of Neutral Current $\pi^{0}$ interactions. Note that the P0D,
like the FGD, has a water target to allow the extraction
of the relevant reaction rates on water.

\section{ND280 TECHNOLOGY CHOICES}

\subsection{Time Projection Chambers}

\begin{figure*}[t]
\centering
\includegraphics[width=60mm]{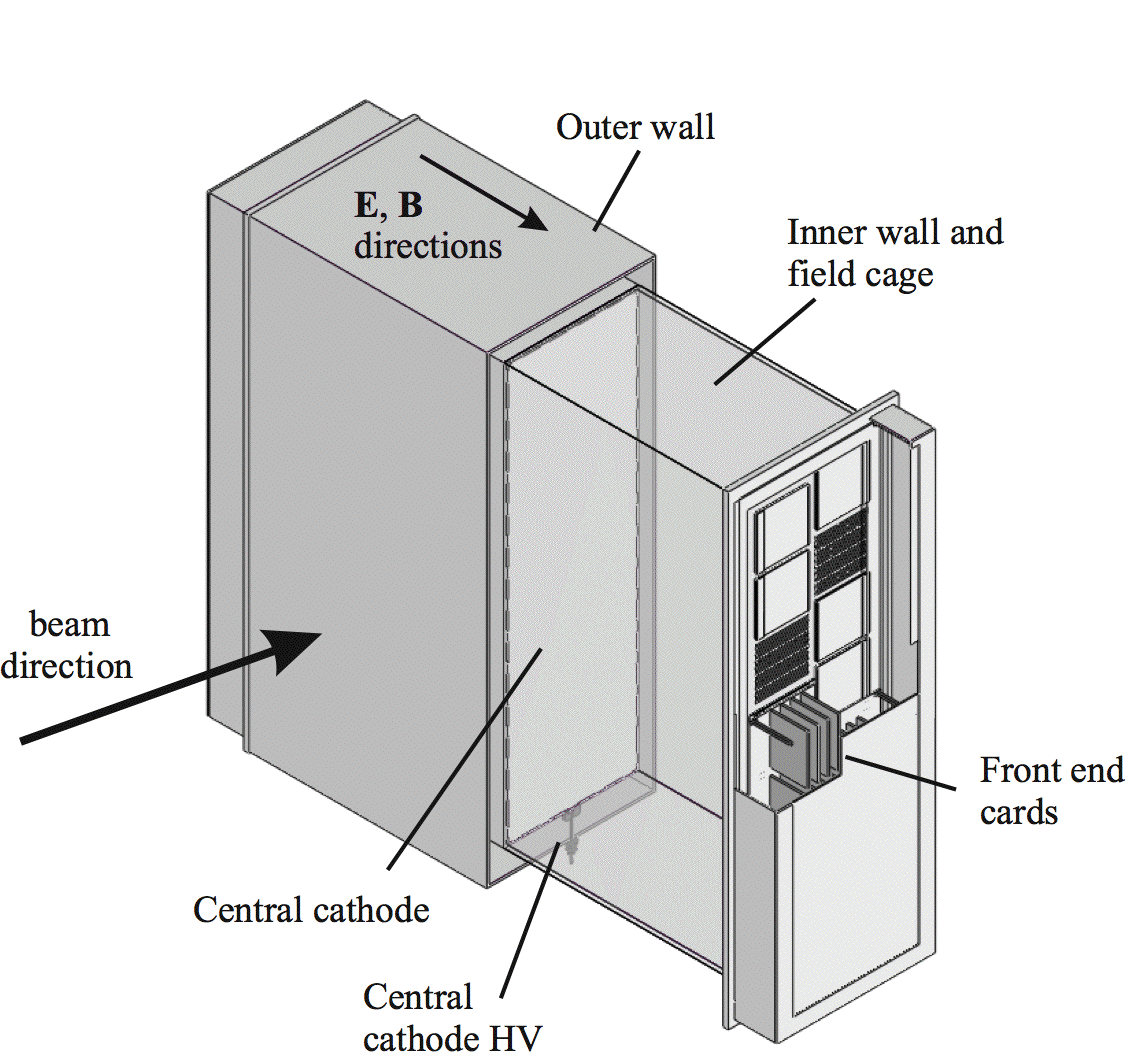}
\includegraphics[width=50mm]{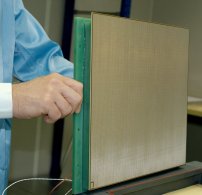}
\caption{Left: overview of the TPC. Right: T2K-TPC $\mu$MEGAS module.} \label{tpc}
\end{figure*}

The three Time Projection Chambers are part of the 
Tracker section of the ND280 detector.  The overview of a single TPC 
is shown on the left side of Figure \ref{tpc}.  Each TPC  
has two inner drift volumes separated by a central cathode.
This is surrounded by an outer box holding CO$_{2}$ gas that serves
as an insulator.

When a charged particle passes through the TPC it will create ionization 
electrons, which will then drift to the readout planes on the sides of the TPC.
Measurement of the drift time, combined with the position of the pad that 
the electrons drift to, allows for excellent 3D reconstruction of each charged particle.
Precise 3D tracking, combined with the 0.2 Tesla magnetic field, results in TPCs
that will achieve momentum resolution of better than 10\% for particles
with momenta below 1 GeV/c.

The T2K TPC uses $\mu$MEGAS modules for the electron amplification and readout.
An example of a  T2K-TPC $\mu$MEGAS is shown on the right side of Figure \ref{tpc}.
The capacity to produce such large $\mu$MEGAS was the result of a significant amount 
of development work.
For electronic readout of the $\mu$MEGAS the TPC group developed a custom ASIC 
that provides a 511 deep switched capacitor array.  The information from this switched
capacitor array is digitized, zero suppressed and then transferred off-magnet 
over a fibre optic link.  The data readout is a challenge, because there are $\approx$100,000 
channels of $\mu$MEGAS data for the full TPC.  This data readout chain was  
copied and used by the FGD sub-detector.

\subsection{Scintillator Sub-Detectors}

Five of the six ND280 sub-detectors use scintillators as the active 
medium.  In general, we use long, thin scintillator bars arranged
into layers; sets of layers with different orientations allows 
for 3D reconstruction of tracks and showers.  Scintillation light 
from the bars is retransmitted by wave-length shifting fibres to photosensors.

\begin{figure*}[t]
\centering
\includegraphics[width=59mm]{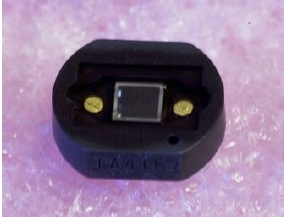}
\includegraphics[width=80mm]{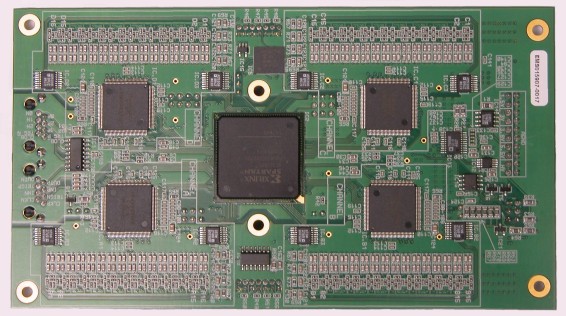}
\caption{Left: a Multi-Pixel Photon Counter. The active area of the MPPC is approximately 1.3 mm$^2$.
Right: a Trip-T front-end board, mounting the ASICs and associated
electronics required for MPPC readout.} \label{scint_elec}
\end{figure*}

The ND280 group has chosen to use Multi-Pixel Photon Counters (MPPCs)
for all of our photosensors ($\approx$50,000 required).  
These particular devices are made by Hamamatsu; but MPPCs
are part of a broader family of devices based on arrays of silicon photodiode pixels.
Each pixel operates above its breakdown voltage (approximately 70V) and can produce an avalanche 
of electrons if struck by a photon.   An MPPC therefore provides the single photon
counting capability and electronic gain of a traditional PMT.  An MPPC is 
superior to a PMT in terms of costs; in addition an MPPC will work well in a magnetic
field, meaning we can place our photosensor right inside our detector.
The left side of Figure \ref{scint_elec} shows an MPPC.
Since MPPCs are new devices, the ND280 group has gone through an extensive period of research
and testing.
We believe that we have a firm understanding of the crucial characteristics of these 
devices, such as the dependence of the breakdown voltage on temperature and their 
long-term stability.

In addition to the photosensors, a great deal of other 
work was involved in the construction of the scintillator based detectors.  In particular,
extensive effort has been made on the development of the electronics readout of the MPPCs.
Most of the sub-detectors  used an ASIC called the Trip-T for electronic readout
(the Trip-T was originally developed at Fermilab for the D0 experiment); this chip 
provides time and charge for any discriminated pulses.  
The right side of Figure \ref{scint_elec} shows an example of one of the front-end boards
that houses these Trip-T chips; in addition to the electronics readout, these 
boards also provide the high voltage to the MPPCs, as well as environment monitoring.
Approximately 1100 of these boards will be produced.

\section{CURRENT STATUS AND CONCLUSIONS}

The ND280 group is making rapid progress towards the construction
and installation of all sub-detector components.
The large pit at JPARC that houses the ND280 facility is 
now complete and work is progressing on installing all the required
services.
Refurbishment and transportation of the UA1 magnet is now complete
and the magnet is installed in the ND280 pit.
By April 2009 the INGRID detector will be installed, 
in time for the first neutrino events provided by JPARC.
All other subdectors are in production at this moment and 
extensive commissioing (including beamline tests) is
proceeding for many groups.
All sub-detectors will be installed at ND280 by Fall 2009.








\end{document}